\documentclass[12pt]{article}
\ifx\pdfoutput\undefined
\usepackage[dvips,bookmarks]{hyperref}	
\else
\usepackage{hyperref}	
\fi
\hypersetup{colorlinks,bookmarksopen,bookmarksnumbered,citecolor=blue,
	pdfstartview=FitH}
\def\hhref#1{\href{http://arxiv.org/abs/hep-th/#1}{hep-th/#1}} 
\def\mhref#1{\href{mailto:#1}{#1}}		

\catcode`@=11
\def\secteqno{\@addtoreset{equation}{section}%
\def\theequation{\thesection.\arabic{equation}}}
\catcode`@=12
\secteqno
\newcommand{\be}{\begin{equation}}
\newcommand{\ee}{\end{equation}}
\newcommand{\bea}{\begin{eqnarray}}
\newcommand{\eea}{\end{eqnarray}}
\newcommand{\bref}[1]{(\ref{#1})}
\newcommand{\nn}{\nonumber}\def\f#1#2{{\textstyle{#1\over#2}}}	   
\def\half{{\textstyle{1\over{\raise.1ex\hbox{$\scriptstyle{2}$}}}}}
\mathcode`\*="702A                  
\catcode162=13 \def\op{\oplus}
\catcode163=13 \def\itm{\relax\ifmmode\to\else\itemize\fi}
\begin{document}
\thispagestyle{empty}
\hfill October 31, 2003

\hfill KEK-TH-925

\hfill YITP-SB-03-56
\vskip 20mm
\begin{center}
{\Large\bf Snyderspace }
\vskip 6mm
\medskip
\vskip 10mm
{\large Machiko\ Hatsuda$^{\ast\dagger}$~and~Warren\ Siegel$^\star$}

\parskip .15in
{\it $^\ast$Theory Division,\ High Energy Accelerator Research Organization (KEK),\\
\ Tsukuba,\ Ibaraki,\ 305-0801, Japan} \\{\it $^\dagger$Urawa University, Saitama \ 336-0974, Japan}\\
{\small e-mail:\ \mhref{mhatsuda@post.kek.jp}} \\
\parskip .35in
{\it 
$^\star$C.N. Yang Institute for Theoretical Physics,\\ 
 State University of New York,\ Stony Brook,\ NY \ 11794-3840, USA}\\
{\small e-mail:\ \mhref{siegel@insti.physics.sunysb.edu} }\\

\medskip
\end{center}
\vskip 10mm
\begin{abstract}
We generalize the construction of Snyder to a Lorentz covariant noncommutative superspace.
\end{abstract} 

\setcounter{page}{1}
\parskip=7pt
\newpage
\section{ Introduction}

Snyder \cite{[1]} introduced noncommutative geometry by considering de Sitter momentum space.  (See also Yang \cite{[2]}.)  Lorentz invariance is preserved, while translation invariance is broken by the finite radius of the momentum space, corresponding to quantization of the conjugate configuration space.  The noncommuting translation operators of this momentum space become the noncommuting position operators of configuration space:
\bea
[ x^a , x^b ] \sim L^{ab}
\eea
in terms of position $x$ and angular momentum $L$.
Gol'fand \cite{[3]} applied this concept to interacting theories.  Here we will review and clarify these results, and then extend them to superspace:
\bea
\{ \theta , \theta \} \sim x + L + T
\eea
in terms of the fermionic coordinates $\theta$ and internal (R) symmetry charges $T$.
This form of noncommutative geometry appeared in a supersymmetric, 
higher-derivative particle theory that reproduces the superstring 
theory prediction of supergravity as one-loop bound states in D=10 \cite{[4]}.

\section{ Bosonic}
\subsection{ Coordinates}

The first example of quantization we learn as students is the particle in a one-dimensional box.  Snyder's basic idea is to consider a box in momentum space, so the coordinates of position space are quantized instead.  The size of the box now acts as an ultraviolet cutoff, rather than an infrared one; such a cutoff is often used in introductory quantum field theory, although no consideration is given to its position space interpretation.  In this case we use periodic boundary conditions; i.e., the box is really a circle, so position takes all negative as well as all positive integer values.  In generalization to higher (four) dimensions, we use a (hyper)sphere, so that Lorentz invariance is preserved.  (We Wick rotate from de Sitter space, so both space and time are quantized.  Also, this is a 4-sphere, rather than the interior of a 3-sphere, used for the usual UV cutoff.)  Clearly each of the four spacetime coordinates is quantized, though they are not simultaneously observable.  (A familiar analog is quantization of angular momentum for the 2-sphere, in terms of polar coordinates.  Here we use Cartesian-style coordinates to manifest Lorentz invariance, and to distinguish angular momentum from position.)

We begin by choosing an appropriate coordinate system.  The simplest choice that manifests SO(D) Lorentz invariance (out of the full SO(D+1) invariance of the D-sphere) is that defined by (azimuthal) stereographic projection (S).  These coordinates cover the entire sphere (if we include ``infinity", as the point on the sphere antipodal to the ``origin"), unlike the (azimuthal) gnomonic projection (G) used by Snyder, which covers only half the sphere, as does the case of (azimuthal) orthographic projection (O).  These three standard ``GSO projections" are obtained by projecting in flat (D+1)-space through the sphere onto a (D-)plane touching it at one point, from a point on the corresponding axis, at the center of the sphere (G), the point on the sphere antipodal to the contiguous point (S), or the point at infinity (O).  The coordinates chosen for the plane are then induced on the sphere.  Map makers conventionally use polar coordinates for the (2-)plane; here we use Cartesian coordinates, to manifest SO(D).  Other Cartesian-like coordinates for the sphere may be useful for other purposes: e.g., equidistant coordinates are natural in the 1D case, where they are just the angle around the circle.

The most general SO(D)-manifest coordinates for the sphere take the form
\bea
\tilde{P}&=& A \left( p - {x\cdot p \over x^2}~ x \right) + {A^2 +x^2 \over A-x^2 A'}
	{x\cdot p \over x^2}~ x \nn\\
g_{mn} &=& {1 \over A^2 +x^2} \left[ \left( \eta_{mn} - {x_m x_n \over x^2} \right)
	+ {(A - x^2 A')^2 \over A^2 + x^2 }{x_m x_n \over x^2} \right] \\
 \sqrt{g}& =& { A-x^2 A' \over (A^2+x^2)^{(D+1)/2} } \nn
\eea
for some function $A(\frac{1}{2}x^2)$ and its derivative $A'$, where $\tilde{P}$ are the translation generators.  The metric is related to the SO(D+1) Casimir
\bea
 {\tilde{P}}^2 + \f12
 (x^{[m}p^{n]})^2 = g^{mn}p_m p_n ~~~.
 \eea
At this point of our discussion $x$ are the coordinates of the sphere. 
 Later we will switch $x$ and $p$; $\tilde{P}$ will then become the position operators.  
 In the stereographic case,
\bea
 A = 1 -\f14 x^2 \Rightarrow \quad \tilde{P} = (1-\f14 x^2)p +\f12x\cdot p~ x,\quad
	g_{mn} = {\eta_{mn} \over (1+\f14 x^2)^2} ~~~.
\eea
This result is identical to the construction of the coset space representation of the ``de Sitter" group SO(D+1) from the coset space representation of the ``conformal" group SO(D+1,1):  The coordinates for the conformal group where $P=p$ directly yield stereographic coordinates for the sphere, with $\tilde{P}= P+K$ in terms of the conformal boosts $K$.

\subsection{ Sums}

We can identify the sphere with the coset space SO(D+1)/SO(D), so a point $p$ on the sphere corresponds to a finite ``translation" $g(p)$ from the origin.  We can then noncommutatively ``add" two points, as defined by performing two consecutive such transformations: 
\bea
g(p)g(q) = g(p\op q)\Lambda
\eea
where the $g$'s are in the coset (translations) while $\Lambda$ is in the subgroup SO(D).  Addition of points is not commutative because it involves parallel displacement of one ``vector" along another in a space with curvature.  
The ambiguity of performing an additional SO(D) transformation is resolved by requiring that points have ``inverses"  under this addition (identically, not just up to an SO(D) transformation):
\bea
g(p)g(\ominus p) = g(\ominus p)g(p) = g(0) = I ~~~.
\eea
We then have the identities
\bea
p\op 0 = 0\op p = p,\quad p\op(\ominus p) = (\ominus p)\op p = 0,
\quad (\ominus p) \op ( p \op q) = q  ~~~.
\eea

The explicit addition rule in a given set of coordinates can be determined easily by examining the case of the 2-sphere, since it involves only 2 points (and the origin).  In particular, for the choice of stereographic coordinates, using complex notation:
\bea
&& g(z) = {1 \over\sqrt{1+z\bar{z}}}\pmatrix{ 1 & -\bar{z} \cr z & 1 \cr },
\quad z = \frac{1}{2}(p_1+ip_2) \label{gz}\\
&\Rightarrow & g(z)g(z') = g(z'')\Lambda,~~z'' = {z+z' \over 1-\bar{z}z'},\quad
	\Lambda= \pmatrix{ e^{ih(z,z')} & 0 \cr 0 & e^{-ih(z,z')} \cr }  ~~~. \label{gz2}
\eea
We then find for stereographic coordinates in arbitrary dimensions
\bea
&& p\op q = 
	{(1-\frac{1}{2}p\cdot q-\f14 q^2)p+(1+\f14 p^2)q \over 1-\frac{1}{2}p\cdot q+\f1{16}p^2 q^2} ~~~.
	\label{gz3}
\eea
(The same result can be obtained directly in higher dimensions by using the corresponding expression for Dirac matrices of SO(D+1).  Gol'fand also considered stereographic coordinates for the transformation, although not for the point itself.)  From this explicit representation we then find
\bea
& \ominus(p\op q) = (\ominus p)\op(\ominus q)  ,\quad 
p\op(q\op r) = 0 \Leftrightarrow  r
\op(q \op p) = 0 &\label{xxsum}
\eea
where in Cartesian-like coordinates $\ominus p=-p$.

Because of the coset nature of this addition, it is also nonassociative:
\bea
&&p\oplus (q\oplus r)=
\left[{c_p~p+c_q~q+c_r~r}\right]/{c_d}\nn\\
&&~~~c_p=
(1-\f1{4}q^2)(1-\f1{4}r^2)-q\cdot r
-\f1{2}(1-\f1{2}q\cdot r-\f1{4}r^2)p\cdot q
-\f1{2}(1+\f1{4}q^2)p\cdot r\nn\\
&&~~~c_q=(1+\f1{4}p^2)(1-\f1{2}q\cdot r-\f1{4}r^2)\nn\\
&&~~~c_r=(1+\f1{4}p^2)(1+\f1{4}q^2)\label{pqr}\\
&&~~~c_d=1-\f1{2}(1-\f1{4}p^2)q\cdot r-\f1{2}(1-\f1{2}q\cdot r-\f1{4}r^2)p\cdot q
-\f1{2}(1+\f1{4}q^2)p\cdot r
\nn\\
&&~~~~~~~~~~~+\f1{16}(p^2q^2+p^2r^2+q^2r^2)\nn
\eea
whereas
\bea
&&(p\oplus q)\oplus r =
\left[{c_p'~p+c_q'~q+c_r'~r}\right]/{c_d'}\nn\\
&&~~~c_p'=
(1-\f1{2}p\cdot q-\f1{4}q^2)(1-\f1{2}s\cdot r-\f1{4}r^2)
\nn\\
&&~~~c_q'=
(1+\f1{4}p^2)(1-\f1{2}s\cdot r-\f1{4}r^2)
\nn\\
&&~~~s=\left\{
(1-\f1{2}p\cdot q-\f1{4}q^2)p+(1+\f1{4}p^2)q
\right\}
/(1-\f1{2}p\cdot q+\f1{16}p^2q^2)
\nn\\
&&~~~c_r'=(1+\f1{4}p^2)(1+\f1{4}q^2)\label{pqrna}\\
&&~~~c_d'=1-\f1{2}(1-\f1{4}r^2)p\cdot q-\f1{2}(1-\f1{2}p\cdot q-\f1{4}q^2)p\cdot r
-\f1{2}(1+\f1{4}p^2)q\cdot r
\nn\\
&&~~~~~~~~~~~+\f1{16}(p^2q^2+p^2r^2+q^2r^2)\nn~~~.
\eea

\subsection{ Products}

The reality condition for a scalar field in momentum space is then
\bea
 \Phi*(p) = \Phi(\ominus p) \label{real}~~~.
 \eea
This leads to the definition of the symmetric inner product, as follows from the usual Hilbert space one:
\bea
 \Phi\cdot\Psi\equiv \int dp~ \Phi(\ominus p) \Psi(p) =\int dp~ \Phi*(p)\Psi(p) 
	=\int dp~ dq~\delta(p\op q)\Phi(p)\Psi(q) \label{inner}
\eea
where $dp$ is the SO(D+1) invariant measure, and similarly for the Dirac $\delta$ function.  (Although we will break SO(D+1) to SO(D), this condition allows the measure to be defined in a coordinate independent way.)

By introducing interactions, Gol'fand effectively introduced an outer product:
\bea
 (\Phi\times \Psi)(\ominus p)\equiv\int dq~dr~\delta [q\op (p\op r)]\Phi(q)\Psi(r) \label{outer}~~~.
 \eea
 From \bref{xxsum}, it follows that this product is commutative (symmetric), in contrast to both the first-quantized product of position operators and the sum of momenta.  In particular,
\bea
 (\Phi\times \Psi)(0) = \Phi\cdot\Psi~~~.
\eea
However, it is not associative:
\bea
(\Xi\times (\Phi\times \Psi))(\ominus p) &=& 
\int dq~dr~ds~\delta \{r \op [(p\op q)\op s]\}\Xi(q)\Phi(r)\Psi(s) ~~, \cr
((\Xi\times \Phi)\times \Psi))(\ominus p) &=& 
\int dq~dr~ds~\delta \{r \op [(p\op s)\op q]\}\Xi(q)\Phi(r)\Psi(s) ~~~.
\eea
Interaction terms can be obtained from the above at $p=0$, or equivalently from expressions such as $\Xi\cdot(\Phi\times\Psi)$.

A simple example is the one-dimensional case, where transformation to position coordinates can be made explicit, due to commutativity of this case.  Choosing the usual angular (equidistant) coordinate $\phi$ for the circle,
\bea
 \tilde{P}= -i\partial_\phi,\quad g_{\phi\phi} = 1 ~~~.
 \eea
We then perform the usual Fourier sum
\bea
 \Phi(\phi) = \displaystyle\sum_{x=-\infty}^\infty \Phi_x e^{i\phi x} \label{Fourier}
 \eea
for integer $x$, the eigenvalue of $\tilde P$.  The inner and outer products are then the obvious ones:
\bea
 \Phi\cdot \Psi= \displaystyle\sum_x  \Phi_x \Psi_x,\quad (\Phi\times\Psi)_x =\Phi_x \Psi_x 
 \eea
Position space then has the obvious discrete symmetry
\bea
 x\itm x+n 
 \eea
for integer $n$.

\section{ D=1}
\subsection{ Superconformal}

We now generalize the bosonic cases discussed in the last section to supersymmetric cases
by considering super-de Sitter momentum space.
One way to derive super-de Sitter coordinates is from 
representations of superconformal group.
(In the next subsection we will also derive them directly.)
There are three methods of derivation of representations of the superconformal group:
by inversion symmetry,
by constraints from a higher-dimensional, linear representation,
or by a coset construction.


The inversion operation is a simple way to 
extend a representation of supersymmetry
to the superconformal group (see, e.g., \cite{1001}). 
In the case of one bosonic dimension,
the conformal boosts $K$ and S-supersymmetry generators $S$
are obtained by the inversion 
\bea
x~\to~-\frac{1}{\lambda^2 x}~~,~~
\theta~\to~-i\frac{\theta}{\lambda x}~~,~~
\bar{\theta}~\to~i\frac{\bar{\theta}}{\lambda x}\label{inversion}
\eea
from the translation generators $P=\partial_x$ and
supersymmetry $Q=\partial_\theta+i\bar\theta\partial_x$.
Under this inversion the supersymmetry-invariant ``metric" 
transforms as
\bea
ds_{\rm flat} = dx+i(d\bar{\theta})\theta+i(d\theta)\bar{\theta}~\to~
\frac{1}{\lambda^2 x^2}~ds_{\rm flat}~~~.
\eea
The one-dimensional de Sitter momentum space, i.e., ${\bf S}^1$, 
has inversion invariance  
in the stereographic coordinate $x$.
In the equidistance coordinate $\phi$,
with $\lambda x=\tan \frac{\phi}{2}$,
the inversion operation is just the constant shift $\phi\to\phi+\pi$. 
 Requiring the inversion invariance under \bref{inversion}
the super-de Sitter metric is given by
\bea
ds_{\rm dS}= {ds_{\rm flat}\over 1+\lambda^2 x^2}
\eea
with the de Sitter space diameter $1/\lambda$.


The superconformal group for D=1 is  OSp(2N$\mid$2).
This OSp(2N$\mid$2) representation can also be derived from manifest 
 OSp(2N$\mid$2) coordinates:
 \begin{enumerate}
\item{ Start with the defining representation of two $x$'s and $2N$ real $\theta$'s.}
\item{ Apply the one constraint 
 $x\cdot p+\theta \cdot \pi=0$ 
 to determine one momentum $p=0$ and gauge fix the corresponding $x=1$.}
  \end{enumerate}
 
Finally, we can also derive superconformal transformations,
directly in finite form, from the ``half-coset" \cite{plc}
 {OSp(2N$\mid$2)/OSp(N$\mid$1)U(1)+}.
 
OSp(2N$\mid$2) is generated by $Q_i, ~S_i$ with $i=1,...,2N$,
and $P,K,\Delta,T_{ij}$, which are generators of the bosonic subgroup
Sp(2)$\otimes$SO(2N):  The algebra is
\bea
&&\left\{Q_i,Q_j\right\}=\delta_{ij}P~~,~~
\left\{S_i,S_j\right\}=\delta_{ij}K~~,~~
\left\{Q_i,S_j\right\}=\delta_{ij}\Delta -\frac{i}{2}T_{ij}\nn\\
&&\left[P,S_i\right]=-iQ_i~~,~~\left[K,Q_i\right]=iS_i~~,~~
\left[\Delta,Q_i\right]=\frac{i}{2}Q_i~~,~~\left[\Delta,S_i\right]=-\frac{i}{2}S_i\nn\\
&&\left[T_{ij},Q_k\right]=-\delta_{k[i}Q_{j]}~~,~~
\left[T_{ij},S_k\right]=-\delta_{k[i}S_{j]}~~\\
&&\left[P,K\right]=-2i\Delta~~,~~
\left[\Delta,P\right]=iP~~,~~
\left[\Delta,K\right]=-iK~~,~~
\left[T_{ij},T_{kl}\right]=\delta_{[l\mid[i}T_{j]\mid k]}~~~.\nn
\eea
This representation of the one-dimensional superconformal generators is
\bea
&Q_i=\pi_i+\frac{1}{2}p\theta_i~~,~~
S_i=xQ_i-\frac{i}{2}\theta_i\theta\cdot\pi&\label{repscon1}
\\
&P=p~~,~~K=x(xp-i\theta \pi )~~,~~\Delta=xp-\frac{i}{2}\theta\cdot \pi~~,~~
T_{ij}=\theta_{[i}\pi_{j]}&\nn
\eea
with $[x,p]=i$ and $\left\{\theta_i,\pi^j\right\}=\delta_i^j$.

\subsection{ Super-de Sitter}

Next we will obtain a representation of the super-de Sitter 
algebra from the superconformal algebra.
Taking linear combinations of generators of OSp(2N$\mid$2),
\bea
\tilde{Q}_a&=&\frac{1}{2}\left\{
(Q_a+iQ_{a+N})-i\lambda (S_a+iS_{a+N})
\right\}\nn\\
\tilde{\bar{Q}}^a&=&\frac{1}{2}\left\{
(Q_a-iQ_{a+N})+i\lambda (S_a-iS_{a+N})
\right\}\label{OSpSU}\\
\tilde{P}&=&\frac{1}{2}(P+\lambda^2 K)\nn\\
\tilde{T}_a{}^b&=&\frac{1}{4}\left\{
(T_{ab}+T_{a+N~b+N})-i(T_{a~b+N}+T_{b~a+N})
\right\}
\eea
with $a=1,...,N$, leads to the following closed algebra:
\bea
&&\left\{\tilde{Q}_a,\tilde{\bar{Q}}^b\right\}=\delta_{a}^{b}
(\tilde{P}-\frac{\lambda}{2N}\tilde{T}_{\rm tracepart})-
\frac{\lambda}{2}\tilde{T}_a{}^b\nn\\
&&\left[\tilde{P},\tilde{Q}_a\right]=-\lambda \tilde{Q}_a~~,~~
\left[\tilde{P},\tilde{\bar{Q}}^a\right]=\lambda \tilde{\bar{Q}}^a\nn\\
&&\left[\tilde{T}_a{}^b,\tilde{Q}_c\right]=\delta_c^b\tilde{Q}_a~~,~~
\left[\tilde{T}_a{}^b,\tilde{\bar{Q}}^c\right]=-\delta_a^c\tilde{\bar{Q}}^b\label{sdS}\\
&&\left[\tilde{T}_a{}^b,\tilde{T}_c{}^d\right]=
\delta_c^b\tilde{T}_a{}^d-\delta_a^d\tilde{T}_c{}^b~~~.\nn
\eea
They generate the subgroup SU(N$\mid$1)
\bea
{\rm OSp(2N}{\mid}2) \supset
{\rm SU(N}{\mid}1)~~~,
\eea
which is the super-de Sitter group.

In terms of the complex coordinates
\bea
\tilde{\theta}^a=\frac{1}{\sqrt{2}}(\theta_a+i\theta_{a+N})~~,~~
\tilde{\bar{\theta}}_a=\frac{1}{\sqrt{2}}(\theta_a-i\theta_{a+N})~~,
\eea
a representation of these super-de Sitter generators is
\bea
\tilde{Q}_a&=&\frac{1}{\sqrt{2}}\left\{
(1+i\lambda x)(\frac{\partial}{\partial \tilde{\theta}^a}+\frac{p}{2}\tilde{\bar{\theta}}_a)
+\frac{\lambda}{2}\tilde{\bar{\theta}}_a(\tilde{\theta}\cdot\frac{\partial}{\partial \tilde{\theta}}
+\tilde{\bar{\theta}}\cdot\frac{\partial}{\partial \tilde{\bar{\theta}}}
 )
\right\}\nn\\
\tilde{\bar{Q}}^a&=&\frac{1}{\sqrt{2}}\left\{
(1-i\lambda x)(\frac{\partial}{\partial \tilde{\bar{\theta}}_a}+\frac{p}{2}\tilde{\theta}^a)
-\frac{\lambda}{2}\tilde{\theta}^a(\tilde{\theta}\cdot\frac{\partial}{\partial \tilde{\theta}}
+\tilde{\bar{\theta}}\cdot\frac{\partial}{\partial \tilde{\bar{\theta}}}
 )
\right\}\label{repscon}
\\
\tilde{P}&=&\frac{1}{2}(1+\lambda^2x^2)p-\frac{i}{2}\lambda^2 x
(\tilde{\theta}\cdot\frac{\partial}{\partial \tilde{\theta}}
+\tilde{\bar{\theta}}\cdot\frac{\partial}{\partial \tilde{\bar{\theta}}}
 )
\nn\\
\tilde{T}_a{}^b&=&-\tilde{\theta}^b\frac{\partial}{\partial \tilde{\theta}^a}
+\tilde{\bar{\theta}}_a\frac{\partial}{\partial \tilde{\bar{\theta}}_b}
\nn~~~.
\eea
The ``super-stereographic projection" 
\bea
\lambda x=\tan \frac{\phi}{2}~~,~~
\tilde{\theta}^a=(1+i\tan \frac{\phi}{2})\vartheta^a~~,~~
\tilde{\bar{\theta}}_a=(1-i\tan \frac{\phi}{2})\bar{\vartheta}_a
\eea
brings the above representation \bref{repscon}
into simpler form as
\bea
\tilde{Q}_a&=&\frac{1}{\sqrt{2}}\left(
\frac{\partial}{\partial \vartheta^a}+\lambda
\bar{\vartheta}_a({-i}\partial_\phi+\bar{\vartheta}\cdot
\frac{\partial}{\partial \bar{\vartheta}})
\right)\nn\\
\tilde{\bar{Q}}^a&=&\frac{1}{\sqrt{2}}\left(
\frac{\partial}{\partial \bar{\vartheta}_a}+\lambda
{\vartheta}^a({-i}\partial_\phi-{\vartheta}\cdot
\frac{\partial}{\partial {\vartheta}})
\right)~~~\\
\tilde{P}&=&\lambda\left({-i}\partial_\phi-\frac{1}{2}
(\vartheta\cdot\frac{\partial}{\partial \vartheta}
-\bar{\vartheta}\cdot\frac{\partial}{\partial \bar{\vartheta}})
\right)\nn\\
\tilde{T}_a{}^b&=&-{\vartheta}^b\frac{\partial}{\partial {\vartheta}^a}
+{\bar{\vartheta}}_a\frac{\partial}{\partial {\bar{\vartheta}}_b}\nn~~~.
\eea
Under an inversion $\phi$ is just constantly shifted,
$\phi\to\phi+\pi$, and the $\vartheta$'s are inert, so
inversion symmetry is manifest in this supercoordinate
system.  It gives consistent
super-Fourier transformation as a natural extension of \bref{Fourier}.

To define inner and outer products \bref{inner} and \bref{outer},
 we need to find a noncommutative addition for supercoordinates, which satisfies \bref{xxsum}, by a construction
 analogous to 
 \bref{gz}, \bref{gz2} and \bref{gz3}. 
It is given by the following parametrization of the coset space
 SU(N${\mid}$1)/SU(N):
 \bea
g(\phi,\theta)=\frac{1}{1+\bar\theta\theta}
\left(
\begin{array}{cc}
e^{i\phi}\left[
(1+\bar\theta\theta)I-2\theta\bar\theta
\right]
&e^{i(N+1)\phi/2}2\theta\\
-e^{i(N+1)\phi/2}2\bar\theta
&
e^{iN\phi}(1-\bar\theta\theta)
\end{array}
\right)~~~,\label{susu}
\eea
where $\theta$ is a column vector and $\bar\theta$ is a row vector.
(The same expression then works for SU(N$|$M)/SU(N)SU(M),
with different normalization of the $\phi$ phase factors.)
The inverse matrix is this matrix with $\phi\to -\phi,~\theta \to -\theta$,
$g(\phi,\theta)^{-1}=g(-\phi,-\theta)$.
The metric is given by
\bea
ds_{\rm dS}
= d\phi\left\{
1-\frac{N-1}{N}\frac{2\bar\theta\theta}{(1+\bar\theta\theta)}
\right\}
+2i\frac{(d\bar{\theta})\theta+(d\theta)\bar{\theta} }{(1+\bar\theta\theta)}~~~.
\eea

As for the bosonic case, the above ``coordinates" $\phi$ and $\theta$
are treated as momenta.  $\tilde P$ becomes the position operator ``$X$",
while $\tilde Q$ becomes the non-anticommutative operator ``$\Theta$"
(not to be confused with the momenta $x$ and $\theta$ used above);
together, $X$ and $\Theta$ are the generalization of the usual
superspace coordinates.  As for the bosonic higher-dimensional case,
where the algebra of coordinate operators closes on angular momentum,
here it closes on the R-symmetry charges.
In units $\lambda=1$, $X$ now takes integer values for bosonic states,
half-integer for fermionic.


\section{ D$>$1}

Here we list the super-(anti-)de Sitter groups 
in all dimensions, and the corresponding superconformal groups: 

\noindent
 \begin{tabular}{|l|l|lll|}
 \hline
{\bf D} &{\bf superconformal} $\to$&{\bf super-de Sitter}
 &{\bf bosonic}&{\bf real}\\
 &&&{\bf subgroup}&{\bf fermions}\\\hline
1 &OSp(2N${\mid}$2)&SU(N${\mid}$1)&SU(N)U(1)  & 2N\\
\hline
2 &OSp(N${\mid}$2)$^2$&OSp(N${\mid}$2)&SO(N)Sp(2)&2N
\\\hline
3 &OSp(N+M$|$4)&OSp(N${\mid}$2)OSp(M$\mid$2)&SO(N)SO(M)Sp(2)$^2$&2(N+M)
\\\hline
4 &SU(N${\mid}$4)&OSp(N${\mid}$4)&SO(N)Sp(4)&4N\\
\hline
5 && SU(N${\mid}$4)&SU(N)SU(4)U(1)&8N\\
\hline
\end{tabular}
\bea
\eea
The coset-space representations of the two groups have the same coordinates:

\noindent
\begin{tabular}{|l|l|l|}
\hline
{\bf D} &{\bf half-coset of superconformal} ~~=&{\bf super-de Sitter coset}\\
 \hline
1 &OSp(2N${\mid}$2)/SO(2N)U(1)+&{SU(N${\mid}$1)/SU(N)}\\
\hline
2 &OSp(N${\mid}$2)$^2$/SO(N)$^2$U(1)$^2$+&OSp(N${\mid}$2)/SO(N)U(1)\\\hline
3 &OSp(N+M$|$4)/SO(N+M)Sp(2)U(1)+&OSp(N${\mid}$2)OSp(M$\mid$2)/SO(N)SO(M)Sp(2) \\\hline
4 &SU(N${\mid}$4)/SU(N)Sp(2)$^2$U(1)$^2$+&{OSp(N${\mid}$4)/SO(N)Sp(2)$^2$}\\\hline
5 && SU(N${\mid}$4)/SU(N)Sp(4)\\
\hline
\end{tabular}
\bea\eea

The one-dimensional super-de Sitter case
was given in the previous subsection \bref{susu}.
The five-dimensional case can be treated in the same way.
For the remaining cases,
the coset OSp(n$\mid$2m)/SO(n) is parametrized by 
\bea
g(\theta,U)=\left(
\begin{array}{cc}
(I-\theta\Omega \theta^T)(I+\theta\Omega\theta^T)^{-1}
&
-2(I+\theta\Omega\theta^T)^{-1}\theta\Omega iU\\
2\theta^T(I+\theta\Omega\theta^T)^{-1}&
(I+\theta^T\theta\Omega)^{-1}(I-\theta^T \theta\Omega)iU
\end{array}
\right)~~
\eea 
where $\Omega$ is the Sp(2m) metric and
\bea
U\Omega U^T\Omega=I~~~.
\eea  
In all 
cases $U$ can be written in the same form as \bref{gz}, where ``$z$" is in R, 
C, H (quaternions) for D = 1, 2, 4.  (For D=4, write the quaternions as 
2$\times$2 complex matrices, and the ``bar" is the quaternion conjugate.)
The inverse of $g$ is obtained just by replacing $\theta\to -\theta$ again
(and the usual for $x$).
The super-de Sitter metric is given by
\bea
ds^2_{\rm dS}={\rm tr}\left[
dUU^{-1}
+2
(I+\theta^T\theta\Omega)^{-1}(\theta^Td\theta-d\theta^T\theta)\Omega
(I+\theta^T\theta\Omega)^{-1} 
\right]^2~~~.
\eea

\section{Conclusions}

We have reviewed Snyder's construction of de Sitter momentum space 
in an explicit way, and extended it to superspace.
Among the ``GSO projection" coordinates,
which are manifestly Lorentz covariant and
give the flat (continuum) limit straightforwardly,
stereographic coordinates are best, since 
they cover the whole space.
The de Sitter coordinates can be obtained 
from representations of the conformal group,
and add noncommutatively.
We have defined the inner and outer products 
of fields on the de Sitter momentum space
by the analog of momentum conservation;
both are commutative, but the outer product
turns out to be 
nonassociative.
``Local" terms in the action can be expressed as the result of
multiple outer products evaluated at 
zero momentum, or with one outer product
replaced with an inner product.

We have extended the bosonic results to super cases.
Super-de Sitter coordinates can be obtained from
representations of superconformal groups.
Three kinds of derivations of representations of superconformal algebras
 have been considered:
by inversion, by constraint from higher dimension, and as coset spaces.
The one-dimensional case lends itself naturally to
``super-stereographic projection", which 
manifests the inversion symmetry of the de Sitter momentum space,
and leads to the super-extension of the equidistance coordinate
used for the definition of Fourier transformation. 

Super-de Sitter coordinates can also be obtained directly as coset spaces,
including representations of finite transformations.
Supercoordinates are given by explicit parametrizaion of the supermatrices, and 
addition of supercoordinates is determined from the group multiplication. 
This parametrization satisfies 
that supervectors have an ``inverse" with respect to the
noncommutative superspace addition (as do the bosonic cases).
Then the super extention of the inner and outer products 
can be defined.

This approach can be useful not only for studies of noncommutative superspace 
(e.g.,  \cite{SchPvN}) but also
to explore supersymmetry in lattice theories.
\par\vskip 6mm
\noindent{\bf Acknowledgments}

 We thank C.N. Yang for discussions on
noncommutative geometry.
M.H. acknowledges the Simons Workshop in Mathematics and Physics 2003,
and
the CNYITP at Stony Brook,
where the main part of this work
has been done, for their hospitality.  
W.S. is supported in part by
National Science Foundation Grant No.\ PHY-0098527.

\end{document}